\documentstyle[12pt,epsfig,amsmath]{article}

\newcommand {\be}{\begin{equation}}
\newcommand {\ee}{\end{equation}}
\newcommand {\ba}{\begin{eqnarray}}
\newcommand {\ea}{\end{eqnarray}}

\begin{document}

\def \a'{\alpha'}
\baselineskip 0.65 cm
\begin{flushright}
\ \today
\end{flushright}

\begin{center}{\large
{\bf Constraints on Randall-Sundrum model from top-antitop
production at the LHC}} {\vskip 0.5 cm} {\bf Seyed Yaser Ayazi and
Mojtaba Mohammadi Najafabadi}{\vskip 0.5 cm } School of Particles
and Accelerators, Institute for Research in Fundamental Sciences
(IPM), P.O. Box 19395-5531, Tehran, Iran
\end{center}

\begin{abstract}
We study the top pair production cross section at the LHC in the
context of Randall-Sundrum model including the Kaluza-Klein (KK)
excited gravitons. It is shown that the recent measurement of the
cross section of this process at the LHC restricts the parameter
space in Randall-Sundrum (RS) model considerably.  We show that
the coupling parameter  ($\frac{k}{\overline{M}_{pl}}$) is
excluded by this measurement from $0.03$ to $0.22$ depending on
the mass of first KK excited graviton ($m_1$). We also study the
effect of KK excitations on the spin correlation of the top
pairs. It is shown that the spin asymmetry in $t\bar{t}$ events
is sensitive to the RS model parameters with a reasonable choice
of model parameters.

\end{abstract}

\section{Introduction}
One of the primary mysteries of high energy physics is the large
difference between scale of the electroweak and fundamental scale
of gravity which is known as the hierarchy problem. One way to
solve the hierarchy problem was proposed by Randall and Sundrum
(RS) in 1999 \cite{Randall:1999ee}. In this model, gravity can
propagate in 5-dimensional non-factorizable geometry and therefore
it generates the 4-dimensional weak Planck scale hierarchy by an
exponential function of the compactification radius which is
called warped factor. In this scenario, an infinite tower of the
KK excited particles appears which their effective couplings with
other particles are characterized by an exponential warped
factor. While in the Arkani-Hamed, Dimopoulos and Dvali (ADD)
model for extra dimension \cite{ArkaniHamed:1998rs}, the mass
difference of each KK excited particle is inversely proportional
to the radius of the extra dimensions and it is more difficult to
detect and identify the KK excited particles at the LHC. As a
consequence, the RS model is more motivated than the ADD model
\cite{ArkaniHamed:1998rs}.

The top quark is the heaviest fundamental particle which have been
discovered to date and might be the first place in which the new
physics effects could appear. The LHC allows us to investigate the
properties of the top quark in details. Very large numbers of top
quarks are produced at the LHC eventually more than $10^7$
$t\overline{t}$ pairs per year \cite{Halkiadakis:2010mj}. This
will make feasible the precise investigations of the top
interactions. At the LHC, top quarks are dominantly produced in
$t\overline{t}$ pairs via the processes $q\overline{q}\rightarrow
t\overline{t}$ and $gg\rightarrow t\overline{t}$.   An intriguing
speciality of the top quark is its extremely short lifetime which
is due to its heaviness. This feature causes that it decays
before it can form any hadronic bound state. As a result, the top
quark spin will be transferred to its decay products.

The cross section value for top pair production have
been measured by CMS experiment recently \cite{CMS}:
\begin{eqnarray}
\sigma(pp\rightarrow t{\overline t} ) & = 165.8\pm13.3~[pb]~(\rm
stat\oplus sys).\label{exp}
\end{eqnarray}

In this paper, we study the effects of RS model on the
$t\overline{t}$ production at the LHC and compare our numerical
results with the experimental measurement.  To study the effects
of RS model, we consider three quantities which are particularly
useful for comparing theory with experimental results. These
quantities are the total cross section of top pair production,
the differential cross section as a function of invariant mass of
$t\bar{t}$, and the spin asymmetry in top-antitop production.

The rest of this paper is organized as follows: In the next section, we summarize the
effects of RS model on the cross section production
of top-antitop and spin asymmetry. In section~3 we compare the
effects of the RS model with the LHC measurements and present our numerical results.
In section 3, we also discuss the constraints on the parameters space of
RS model using the future measurements of spin asymmetry.
The conclusions are given in section~4.

\section{The effects of RS model on $t\bar{t}$ production}

In this section, we briefly describe the model and study the effects
of KK excited gravitons on top-antitop production at the LHC. The
set up of RS model consists of two D3-branes which embedded in a
five dimensional bulk spacetime \cite{Randall:1999ee}. This
theory has been considered a 5-dimensional non-factorizable
metric \cite{Randall:1999ee}:
\begin{eqnarray}
ds^2=e^{-2k r_c\pi}\eta_{\mu\nu}dx^{\mu}dx^{\nu}-r_c^2d\phi^2,
\end{eqnarray}
where $x_{\mu}$ are the coordinates for the familiar four dimensions,
$\phi$ ($0\leq\phi\leq\pi$) is the coordinate for an extra
dimension and $r_c$ is a compactification radius. Two branes are
localized at $\phi=0$ and $\pi$ which are called hidden and
visible branes, respectively. Here $k$ is the $\rm AdS$ curvature
scale which is of order the Planck scale.

Effective Lagrangian in 4-dimensional effective theory is given by
\cite{Davoudiasl:1999jd}:
\begin{eqnarray}
{\cal{L}}_{int}&=&
-\frac{1}{M^{3/2}}T^{\alpha\beta}(x)h_{\alpha\beta}(x,\phi=\pi),
\end{eqnarray}
where $M$ is the known Planck scale, $T_{\alpha\beta}$ is the symmetric energy-momentum tensor of the
SM fields on the visible brane and $h_{\alpha\beta}$ are graviton
fields. After expanding the graviton field into the KK states,
Lagrangian is found as:
\begin{eqnarray}
{\cal{L}}_{int}&=&
-\frac{1}{\overline{M}_{pl}}T^{\alpha\beta}(x)h^{(0)}_{\alpha\beta}-\frac{1}{\Lambda{_{\pi}}}T^{\alpha\beta}(x)\sum_{n=1}^{\infty}h^{(n)}_{\alpha\beta}(x),
\label{inter}
\end{eqnarray}
where $\overline{M}_{pl}$ is the reduced Planck mass. Examining
of the RS action in the 4-D effective theory yields the relation:
\begin{eqnarray}
\overline{M}^2_{Pl}=\frac{{M}^3}{k}(1-e^{-2k r_c\pi}).
\end{eqnarray}
To solve the hierarchy problem, it is assumed that gravity is
localized on the brane at $\phi=0$ and $k r_c\simeq11-12$
\cite{Davoudiasl:1999jd}. In this situation, $\rm TeV$ scale
theory can naturally be attained on the 3-brane at $\phi=\pi$.
The scale of physical processes on this $\rm TeV$-brane is then
$\Lambda_{\pi}\equiv \overline{M}_{pl}e^{-k r_c\pi}$. According
to Eq. \ref{inter}, the graviton zero-mode couples to the SM
fields with usual strength while the KK modes strongly couples to
the SM fields with the suppression factor of $1/\Lambda_{\pi}$.

The masses of the KK excitations of graviton are given by:
\begin{eqnarray}
m_n & = x_n k e^{-k r_c\pi}, \
\end{eqnarray}
where $x_n$ is a root of the Bessel function of the first order.
In this paper, we study the effect of exchanged KK excited
gravitons on the $t\bar{t}$ production cross section. The leading
order (LO) processes for the production of top pair at the LHC
include these two processes $q\overline{q}\rightarrow
t\overline{t}$ and $gg\rightarrow t\overline{t}$. The
corresponding invariant matrix elements for these processes
including the KK graviton modes effects have been calculated in
\cite{Arai:2007ts,hep-ph/0701150,hep-ph/0010010}. When
$q\overline{q}$ is the initial state, the amplitude for the same
spin direction of top-antitop and opposite spin direction are
given by:
\begin{eqnarray}
|{\cal{M}}(q\overline{q}\rightarrow
t_\uparrow\overline{t}_\uparrow)|^2&=&|{\cal{M}}(q\overline{q}\rightarrow
t_\downarrow\overline{t}_\downarrow)|^2
\nonumber\\
&=&\frac{g^4}{9}(1-\beta^2)\sin^2\theta+\frac{|A(s)|^2s^4\beta^2}{128}(1-\beta^2)\sin2\theta,
\label{amp1}
\end{eqnarray}
\begin{eqnarray}
|{\cal{M}}(q\overline{q}\rightarrow
t_\uparrow\overline{t}_\downarrow)|^2&=&|{\cal{M}}(q\overline{q}\rightarrow
t_\downarrow\overline{t}_\uparrow)|^2
\nonumber\\
&=&\frac{g^4}{9}(1+\cos^2\theta)+\frac{|A(s)|^2s^4\beta^2}{128}(\cos^22\theta+\cos^2\theta),
\end{eqnarray}
where $\beta=\sqrt{1-4m_t^2/s}$ and $m_{t}$ is the top quark mass. Here
$A(s)$ is defined by
\begin{eqnarray}
A(s)=-\frac{1}{\Lambda^2_\pi}\sum_{n=1}^{\infty}\frac{1}{s^2-m^2_n+im_n\Gamma_n},
\end{eqnarray}
where $s$ is partonic centre-of-mass energy and $\Gamma_n$ is the
total decay width of the $n$-th KK graviton excitation
\cite{Arai:2007ts}. In this paper, we take to account the effects
of KK excited gravitons up to $n=5$. For larger values of $m_n$,
the effects of KK excitations on observables are negligible. In
Eq. \ref{amp1}, $\theta$ is the angle between the incoming  and
outgoing top quark, $g$ is the strong coupling constant. For the
situation with gluon-gluon  in the initial states, the amplitude
is given by:
\begin{eqnarray}
|{\cal{M}}(gg\rightarrow
t_\uparrow\overline{t}_\uparrow)|^2&=&|{\cal{M}}(gg\rightarrow
t_\downarrow\overline{t}_\downarrow)|^2 \nonumber\\
&=&\frac{g^4\beta^2}{96}{\cal{D}}(\beta,\cos\theta)(1-\beta^2)(1+\beta^2+\beta^2\sin^4\theta)\nonumber\\
&+&{\cal{Z}}(\beta,\theta,s)s^2\beta^2(1-\beta^2)\sin^4\theta,
\end{eqnarray}
\begin{eqnarray}
|{\cal{M}}(gg\rightarrow
t_\uparrow\overline{t}_\downarrow)|^2&=&|{\cal{M}}(gg\rightarrow
t_\downarrow\overline{t}_\uparrow)|^2 \nonumber\\
&=&\frac{g^4\beta^2}{96}{\cal{D}}(\beta,\cos\theta)\sin^2\theta(1+\cos^2\theta)\nonumber\\
&+&{\cal{Z}}(\beta,\theta,s)s^2\beta^2\sin^2\theta(1+\cos^2\theta).
\end{eqnarray}
In the above Equations,
${\cal{D}}(\beta,\cos\theta)$ and ${\cal{Z}}(\beta,\theta,s)$ are
defined by:
\begin{eqnarray}
{\cal{D}}(\beta,\cos\theta)=\frac{7+9\beta^2\cos^2\theta}{(1-\beta^2\cos^2\theta)^2},
\end{eqnarray}
\begin{eqnarray}
{\cal{Z}}(\beta,\theta,s)=\frac{1}{32}(-\frac{g^2}{1-\beta^2\cos^2\theta}\rm
Re(A(s))+\frac{3}{8}|A(s)|^2s^2).
\end{eqnarray}

The total cross section for production of $t\bar{t}$ with spin indices $\alpha$ and $\beta$ has the following form:
\begin{eqnarray}
\sigma(pp\rightarrow t_{\alpha}{\overline t}_{\beta}) & =
&\sum_{ab} \int dx_1dx_2f_a(x_1,Q^2)f_b(x_2,Q^2)
\widehat{\sigma}(ab\rightarrow t_{\alpha}{\overline t}_{\beta}), \
\end{eqnarray}
where $f_{a,b}(x_i,Q^2)$ are the parton structure functions of proton.
$x_1$ and $x_2$ are the parton momentum fractions and $Q$ is the factorization scale.

\section{Numerical results}

In the previous section, we introduced the total cross section of
top pair production at the LHC. In this section, we calculate the differential cross section and spin asymmetry and compare them
with the ongoing measurement at the LHC.

We are also interested in the differential cross section as a
function of invariant mass
$M_{t\overline{t}}=\sqrt{(p_t+p_{\overline{t}})^2}$, where $p_t$
and $p_{\overline{t}}$ are the four-momenta of top and anti-top,
respectively. This quantity is defined by

\begin{eqnarray}
\frac{d\sigma(pp\rightarrow
t_{\alpha}\overline{t}_{\beta})}{dM_{t\overline{t}}}=\sum_{ab}\int_{\frac{M^2_{t\overline{t}}}{E^2_{CMS}}}^1dx_1\frac{2M^2_{t\overline{t}}}{x_1E^2_{CMS}}f_a(x_1,Q^2)
f_b(\frac{M^2_{t\overline{t}}}{x_1E^2_{CMS}},Q^2)\widehat{\sigma}(ab\rightarrow
t_{\alpha}{\overline t}_{\beta}).
\end{eqnarray}

The best way to study the top pair spin correlation is to analyze
the angular correlations of two charged leptons originating from
the full leptonic top-antitop decay. Spin asymmetry between the
top-antitop pair is defined as \cite{Brandenburg:1996df}:
\begin{eqnarray}
A=\frac{\sigma_{\uparrow\uparrow}+\sigma_{\downarrow\downarrow}-\sigma_{\uparrow\downarrow}-\sigma_{\downarrow\uparrow}}
{\sigma_{\uparrow\uparrow}+\sigma_{\downarrow\downarrow}+\sigma_{\uparrow\downarrow}+\sigma_{\downarrow\uparrow}}\label{delta}
\end{eqnarray}
where $\sigma_{\uparrow/\downarrow\uparrow/\downarrow}$ are the
production cross section for top pair  with  characterized spin
indices. In the context of SM, spin asymmetry has been calculated
at $7~\rm TeV$ center-of-mass energy to be $0.16$ for the LHC while it is $0.32$ for
$14~\rm TeV$. The CMS collaboration hopes to measure spin
asymmetry of this process by several percent \cite{CMS
performance}.

In the rest of this section, we study the effects of KK excitation
of gravitons on the production cross section of  $pp\rightarrow
t\overline{t}$ at the LHC. We first discuss the bounds on
parameters space of the RS model then we present our results and
discuss the possibility of restricting the RS model parameters at
the LHC via top pair production observables.

In the RS model, there are two main parameters which are the first
excitation graviton mass ($m_1$) and the coupling parameter
($\frac{k}{\overline{M}_{pl}}$). There are two kinds of
experimental constraints on the parameters of RS model. First
kind of constraints arises from indirect searches for the large
extra dimensions. Model dependent limits can be placed on the
masses of KK excitation fields from the precision electroweak
tests \cite{precisin test}, cosmology ($e.g$., expansion rate of
the universe) \cite{cosmology}, black hole production at colliders
\cite{blachhole}, flavor observables and Higgs related collider
searches \cite{flavorobs}, and low energy experiments ($e.g.$, CP
violation and FCNC processes) \cite{CP RS}. Another kind of
constraints comes from the direct searches for KK gravitons at
Tevatron \cite{Tevatron} and LHC
\cite{directphoton,directdielectron}. In \cite{directphoton}, RS
gravitons decaying to pairs of photons have been studied  and
shown that for values of the coupling parameter
($\frac{k}{\overline{M}_{pl}}$) ranging from 0.01 to 0.1, at
$95\%$ confidence level, the lower bounds on excited graviton
masses vary from than 371 to $945~\rm GeV$. Paper
\cite{directdielectron} describes a search for high mass
resonance decaying to electron pairs. It is shown that the
corresponding limits on RS graviton production for coupling of
$0.1$ and $0.05$ are 930 and $730~\rm GeV$, respectively.
\begin{figure}
\centerline{\includegraphics[scale=0.6]{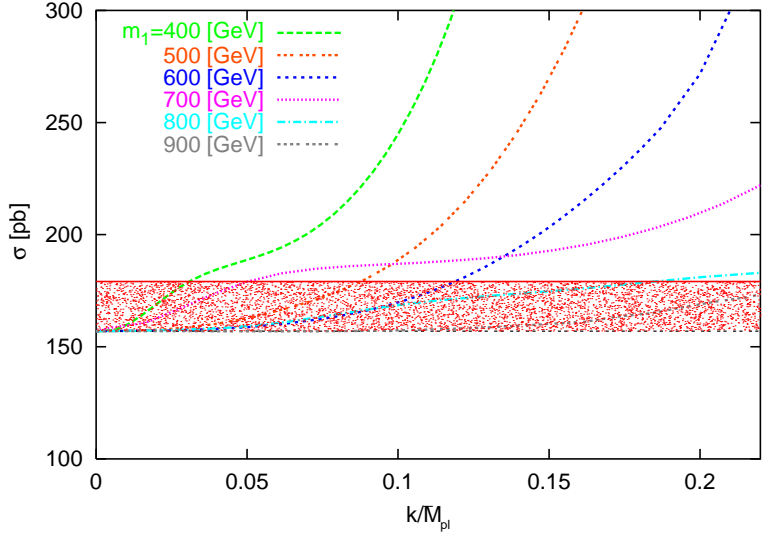}} \caption{The
total cross section of $t\bar{t}$ production at the LHC as a
function of $\frac{k}{\overline{M}_{pl}}$. The horizontal dotted
black line at 157 $~\rm pb$ depicts the theoretical SM value and
dotted red line at $179.1 ~\rm pb$ depicts the present upper
experimental limit on the cross section.}\label{cross}
\end{figure}

In this paper, to calculate $\sigma(pp\rightarrow t\overline{t})$,
we have used the MSTW parton structure functions \cite{MSTW} and
set the center-of-mass energy to $7~ \rm TeV$. To study the
effects of KK excited gravitons on top-antitop production, we
display the cross section of $pp\rightarrow t\overline{t}$ versus
$\frac{k}{\overline{M}_{pl}}$ in Figure-\ref{cross}. In
Fig.~\ref{cross}, we have considered different values for $m_1$.
The horizontal dotted black line at $157$ $\rm pb$ depicts the
theoretical SM prediction and the dotted red line at $179.1$ $\rm
pb$ depicts the present upper experimental limit on the cross
section.  Obviously, all curves asymptotically tend to the SM
prediction when $\frac{k}{\overline{M}_{pl}} \rightarrow 0$.
Fig.~\ref{cross} shows that for the region of
$\sigma(pp\rightarrow t\overline{t})$ above the present
experimental bound, coupling parameter
($\frac{k}{\overline{M}_{pl}}$) is excluded. Hatched area shows
values for $\frac{k}{\overline{M}_{pl}}$ which have not been
excluded by the present upper experimental limit. Notice that for
various values of $m_1$, bounds on $\frac{k}{\overline{M}_{pl}}$
are different but this dependence is not linear. As it can be
seen in Fig.\ref{cross}, for excited graviton masses of $m_1=$
400, 500, 600, 700, 800 and $900~\rm GeV$, the values of the
coupling parameter ($\frac{k}{\overline{M}_{pl}}$) must be
smaller than 0.02, 0.06, 0.1, 0.035, 0.11 and 0.2, respectively.
The limits are stronger than some of the bounds coming from other
approaches \cite{directphoton,directdielectron}.

\begin{figure}
\centerline{\includegraphics[scale=0.5]{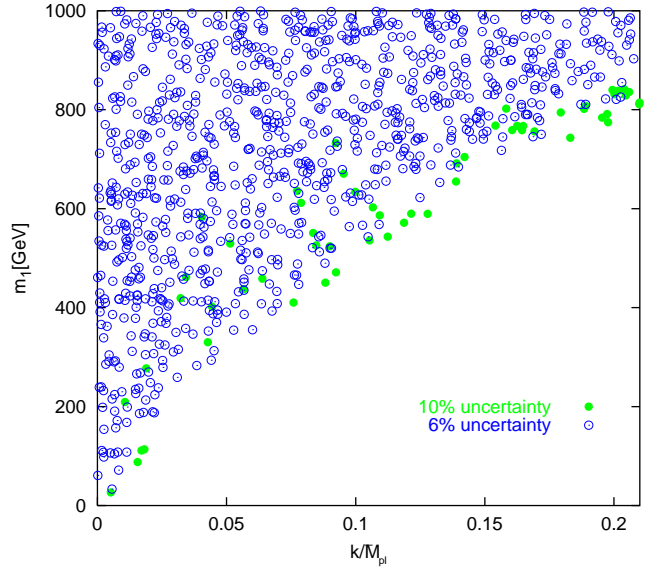}}
\caption{Scatter plot of $m_1$ versus
$\frac{k}{\overline{M}_{pl}}$. Plot is showing the points (filled
and empty) for which the total cross section of top-antitop
production are in the range of experimental uncertainty as
mentioned in relation (\ref{exp}). The points depicted by filled
circle (green) are the region where the total cross section of
top-antitop production are in the range of the present precision
experimental limit with $10\%$ uncertainty. The points depicted
by empty circle (blue) show the points at which the total cross
section of top-antitop production is in the range that can be
probed in near future with $6\%$ uncertainty.}\label{Scater1}
\end{figure}

To illustrate this observation, we have shown a scatter plot in
Fig.\ref{Scater1}. This diagram depicts scatter plot of $m_1$
versus $\frac{k}{\overline{M}_{pl}}$. Fig.\ref{Scater1} is showing
the points (filled and empty) for which the total cross section
of top-antitop production are in the range of experimental
uncertainty as mentioned in relation (\ref{exp}). For drawing this
plot, we have selected various random values for $m_1$ and
$\frac{k}{\overline{M}_{pl}}$. At points marked by filled circle
(green), the total cross section of top-antitop production
corresponds to the present precision of the experimental
measurement in relation (\ref{exp}). The scatter points depicted
by empty circles (blue) are the points that the total cross
section of top-antitop production correspond to the range that
can be probed in near future with $6\%$ uncertainty. If we assume
to measure the top pair cross section with $6\%$ uncertainty, the
filled circle will be excluded. The future experimental
uncertainty of the LHC on the cross section measurement of
top-antitop provides slightly better constraints on the value of
on the value of $\frac{k}{\overline{M}_{pl}}$, as shown in also
shows that by increasing value of  the $m_1$, constraints on
value of the $m_1$, constraints on $\frac{k}{\overline{M}_{pl}}$
will

\begin{figure}
\centerline{\includegraphics[scale=0.5]{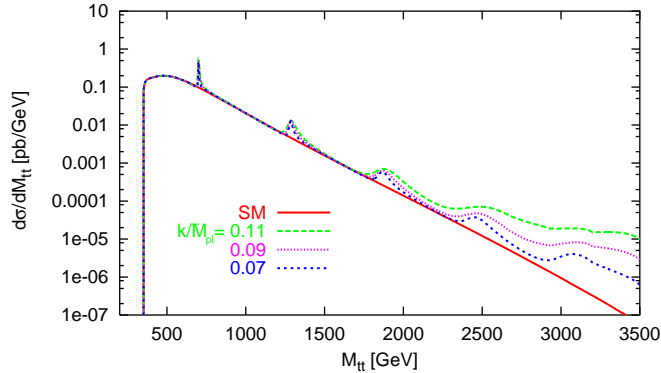}}
\caption{Differential cross section of top-antitop production at
the LHC as a function of invariant mass $M_{t\overline{t}}$. We
have set $m_1=700$~GeV and considered different values for
$\frac{k}{\overline{M}_{pl}}$. The red curve, dashed (green)
curve, small dotted (pink) curve and large dotted (blue) curve
respectively correspond to SM, and RS model with
$\frac{k}{\overline{M}_{pl}}=$ 0.11, 0.09 and 0.07.} \label{dM}
\end{figure}

In Fig. \ref{dM}, we have displayed  the differential cross
section of top-antitop production at LHC as a function of
invariant mass of top pair $M_{t\bar{t}}$. In this figure,
$m_1=700$~GeV and considered different values for
$\frac{k}{\overline{M}_{pl}}$. The red curve, dashed (green)
curve, small dotted (pink) curve and large dotted (blue) curve
correspond to the SM prediction, and the RS model with
$\frac{k}{\overline{M}_{pl}}=$ 0.11, 0.09 and 0.07, respectively.
As it is shown, there are large deviations around the masses of
KK excited garvitons in the RS model from the SM prediction. So
far, $t\overline{t}$ invariant mass distribution has not been
analyzed to search for RS model. Similar analysis for SM-like
$Z'$ has been performed at the LHC \cite{invatriant mass} but
with dielectron final state. In \cite{invatriant mass}, no
significant deviation from the SM expectation has been observed.
\begin{figure}
\centerline{\includegraphics[scale=0.6]{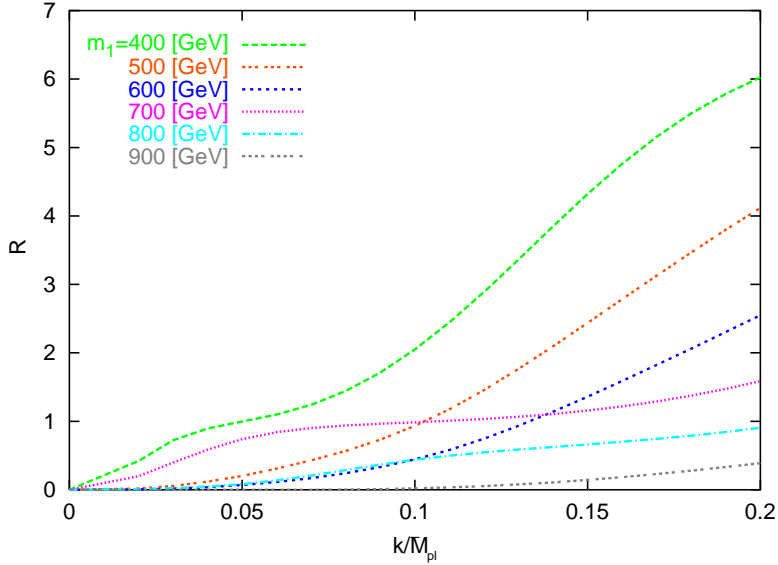}} \caption{
Relative change in spin asymmetry of top-antitop production at the
LHC as a function of $\frac{k}{\overline{M}_{pl}}$. We have
considered different values of $m_1$.}\label{deltaA}
\end{figure}

The top quark spin asymmetry in dilepton channel at the LHC is
predicted with a small uncertainty in the SM \cite{CMS
performance}. For this reason, the spin correlation in $t\bar{t}$
production at the LHC is a good way to search for beyond SM. For
a better study of RS model, we consider the relative change in
spin asymmetry which is defind as:
\begin{eqnarray}
 R=\frac{A-A_{SM}}{A_{SM}}, \
\end{eqnarray}
where $A$ is spin asymmetry in the presence of KK excited
gravitons and $A_{SM}$ is the SM prediction for spin asymmetry.
The relative change in spin asymmetry of the top-antitop
production at the LHC as a function of
$\frac{k}{\overline{M}_{pl}}$ is shown in Fig.~\ref{deltaA}. In
this figure, we have considered different values of $m_1$.

In \cite{Hubaut:2005er}, it has been shown that a $4\%$ precision
on spin asymmetry measurement of $t\overline{t}$ production at
the LHC is possible with $10~\rm fb^{-1}$ . As it can be seen in
Fig.~\ref{deltaA}, deviations from SM prediction in RS model can
be large depending on the model parameters. Therefore,
measurement of spin asymmetry in $t\overline{t}$ pair production
can  help us probe new physics effects.
\begin{figure}
\centerline{\includegraphics[scale=0.5]{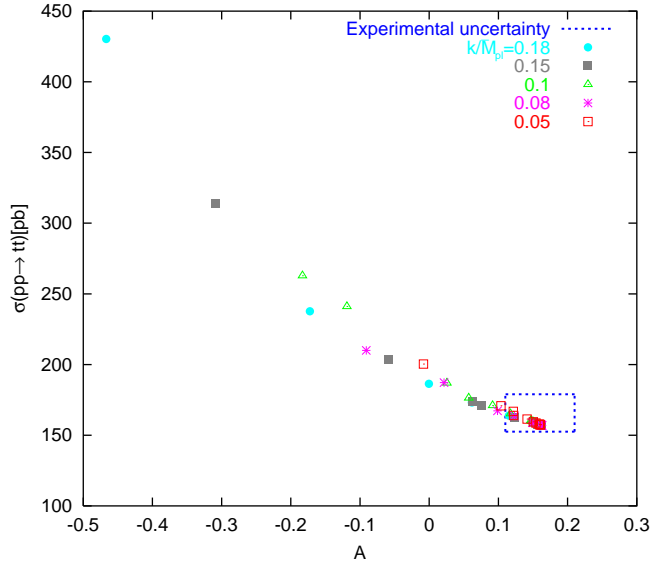}}
\caption{Scatter plot of the total cross section of
$t\overline{t}$ pair production at the LHC versus spin
asymmetry.}\label{Scater2}
\end{figure}

Fig.~\ref{Scater2} is the scatter plot of $t\bar{t}$ production
cross section at the LHC versus the spin asymmetry. This plot is
showing different values for the total cross section of
top-antitop and spin asymmetry in situation that various inputs
have been selected  for $m_1$ and $\frac{k}{\overline{M}_{pl}}$.
The rectangle shows the present experimental uncertainty for the
total cross section (relation \ref{exp}) and the possible
precision of the spin asymmetry measurement with the SM
predictions in center \cite{CMS performance}. As it can be seen in
Fig.~\ref{Scater2}, there are several points which their total
cross sections and spin asymmetry do not violate experimental
bounds and therefore there is a hope to observe RS effects at the
LHC.
\begin{figure}
\centerline{\includegraphics[scale=0.5]{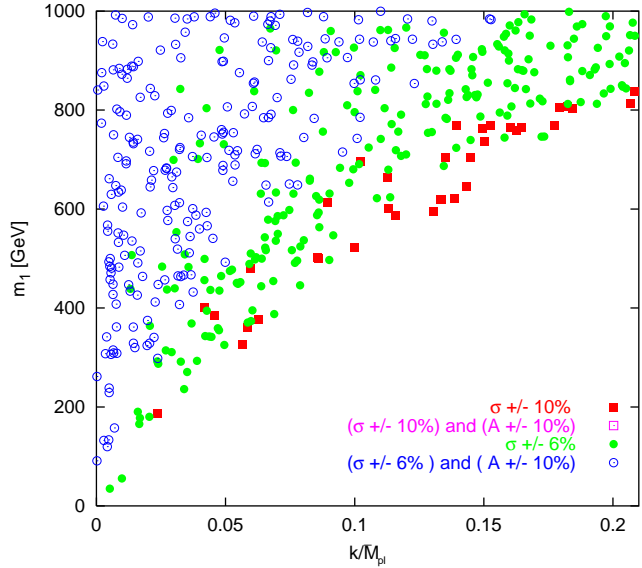}}
\caption{Scatter plot of $m_1$ versus
$\frac{k}{\overline{M}_{pl}}$. Scatter plot is showing the points
(filled and empty) for which the total cross section of top pair
production correspond to the present and future experimental
uncertainties without and with including possible spin asymmetry
measurement around the SM prediction. }\label{Scater3}
\end{figure}

In order to obtain stronger bounds on $m_1$ and
$\frac{k}{\overline{M}_{pl}}$, we have simultaneously considered
cross section and spin asymmetry in $t\overline{t}$ pair
production in Fig.~\ref{Scater3}. In Fig.~\ref{Scater3}, we have
drawn a scatter plot of $m_1$ versus
$\frac{k}{\overline{M}_{pl}}$. The scatter plot is showing points
(filled and empty) for which the total cross section of top pair
production correspond to the experimental uncertainty measurement
as it is mentioned in relation \ref{exp} and spin asymmetry is
within the future uncertainty of $10\%$. Notice that $m_{1}$ and
$\frac{k}{\overline{M}_{pl}}$ have been generated randomly. The
empty squares (pink) in Fig. \ref{Scater3} are the points at
which the total cross section of top-antitop production
corresponds to the experimental uncertainty as mentioned in
relation (\ref{exp}), and the spin asymmetry is in the range of
SM and the spin asymmetry is in the range of SM prediction with
point which can satisfy these conditions. This means that by
measuring spin asymmetry with $10\%$ uncertainty, we can derive
new bounds on RS parameters. The filled squares (red) show
similar situation except that we relax any constraints coming
from spin asymmetry. The points depicted by empty circles (blue)
are the points at which the total cross section of top-antitop
production is in the range which can be probed in near future
with $6\%$ uncertainty and the spin asymmetry is in the range of
SM prediction with $10\%$ uncertainty. If we assume to measure
the cross section of $t\overline{t}$ with $6\%$ uncertainty, the
filled squares will be excluded. The filled circles (green) are
similar to blue circles except that we relax any constraints
coming from spin asymmetry. This figure shows, with simultaneous
measurement of the total cross section and spin asymmetry of
top-antitop production at the LHC, we can derive stronger bounds
on $\frac{k}{\overline{M}_{pl}}$ in comparison with using single
measurement of total cross section. It also can be seen that for
large values of $m_1$, limits are weaker.

\section{Concluding remarks}
In this paper, we have studied the effects of KK excited gravitons
on $t\bar{t}$ production cross section at the LHC with the
center-of-mass energy of $\sqrt{s}=7~ \rm TeV$. We have shown
that the contribution of this model to the cross section
production of $t\overline{t}$ can exceed the present experimental
measurement. From this observation the parameter space of RS
model is restricted. We have shown that for the excited graviton
masses $m_1=$ 400, 500, 600, 700, 800 and $900~\rm GeV$, the
values of the coupling parameter ($\frac{k}{\overline{M}_{pl}}$)
must be smaller than 0.02, 0.06, 0.1, 0.035, 0.11 and 0.2,
respectively. Also we showed that for various values of $m_1$,
restriction on restriction on $\frac{k}{\overline{M}_{pl}}$ can be
different but this completely linear and with the growth of
$m_1$, it does not decrease in all regions. The limits on RS
parameters which arise from studying of production cross section
of top-antitop can be stronger than the ones obtained from other
approaches.

 Assuming a relative uncertainty of $6\%$ on the
measurement of cross section of top-antitop in near future
measurement at LHC, We have shown $\frac{k}{\overline{M}_{pl}}$
can slightly be constrained. We discussed that using the spin
asymmetry $A$, one can derive stronger bound on coupling parameter
($\frac{k}{\overline{M}_{pl}}$).

\end{document}